# Data-driven Design of Isotropic and High-Stiffness TPMS-based Amorphousness-Induced Architected Material (TAAM)

Minwoo Park[a,+], Junheui Jo[a,+], and Seunghwa Ryu[a,*]

[a]*Department of Mechanical Engineering, Korea Advanced Institute of Science and Technology, 291 Daehak-ro, Yuseong-Gu, Daejeon 34141, Republic of Korea*

**Abstract**

For their excellent stiffness-to-weight characteristics, triply periodic minimal surfaces (TPMS) are widely adopted in architected materials. However, their geometric regularity often leads to elastic anisotropy, limiting their effectiveness under complex loading. To address this, we propose TPMS-based amorphousness-induced architected materials (TAAMs), which incorporate controllable geometric disorder as a tunable design variable. This concept of *designable amorphousness* broadens the geometric design space, enabling the simultaneous optimization of stiffness and isotropy. A data-driven framework integrating computational homogenization with multi-objective Bayesian optimization is employed to discover high-performance TAAMs. Selected designs were fabricated using fused deposition modeling and validated through uniaxial compression tests, showing strong agreement with numerical predictions. Compared to conventional TPMS, TAAMs exhibit significantly improved elastic isotropy while maintaining high stiffness across various relative densities. This approach offers a robust and scalable pathway for developing architected materials tailored to applications requiring isotropic performance, such as biomedical implants, protective systems, and aerospace components.

[+]These authors contributed equally: Minwoo Park, Junheui Jo

[*]Corresponding author e-mail: ryush@kaist.ac.kr

# 1. Introduction

Recent advances in mechanical metamaterials have enabled the development of architected materials with extraordinary mechanical properties that surpass those of conventional bulk materials [1-3]. By precisely engineering their internal microarchitectures, these materials can achieve high specific stiffness, superior energy absorption, and unusual behaviors such as negative Poisson's ratio [4-6]. These attributes make them highly suitable for lightweight structural applications, energy-absorbing devices, and biomedical implants. The rapid progress of additive manufacturing (AM) further facilitates the realization of complex microstructures with high geometric fidelity, accelerating research into multifunctional architected materials [7].

Among various design approaches, triply periodic minimal surface (TPMS)-based structures have gained widespread interest due to their smooth and continuous open-cell morphologies, which enable high stiffness-to-weight ratios and alleviate local stress concentrations [8-13]. Their geometry can be precisely modulated by mathematical parameters, enabling tailored control over mechanical performance via relative density and spatial configurations [13, 14]. In particular, the compatibility of TPMS structures with AM techniques such as fused deposition modeling (FDM) or vat photopolymerization has enabled direct fabrication without internal supports, a crucial advantage for design-to-manufacture workflows. However, despite these advantages, their application remains constrained by a fundamental limitation—elastic anisotropy—stemming from their periodic and symmetric geometry [15-17]. This direction-dependent stiffness leads to reduced reliability under complex or uncertain loading conditions, which is critical in fields such as biomedical implants and crash protection systems [18-20].

Various strategies have been explored to mitigate this anisotropy inherent in TPMS-based architected materials [21-27]. One approach involves union operations between multiple TPMS types to generate composite geometries with improved uniformity; however, such operations frequently compromise the defining surface continuity of TPMS structures [25]. Hybridization strategies that combine dissimilar TPMS unit cells have also been examined, but these often suffer from poor intercell connectivity, particularly at low relative densities [26]. Another line of research draws inspiration from crystal lattice symmetries such as simple cubic (SC) and face-centered cubic (FCC) arrangements [28] to impose directional balance within the structure [27]. This approach has been extended through level-set-based mid-surface offsetting techniques to further reduce residual anisotropy [29, 30]. In addition, variable thickness modulation, achieved by parametric control over surface gradients, has been proposed to tailor the directional distribution of stiffness [24].

Recent works have achieved excellent mechanical performance using data-driven frameworks with HTAM [31] and CFCC [32], demonstrating the potential of advanced architected materials. However, these designs typically operate within narrow structural subspaces, either via motif hybridization or symmetry-constrained assembles, limits their flexibility in exploring broader stiffness–isotropy trade-offs. Consequently, achieving an optimal balance between stiffness and isotropy remains a significant challenge in the design of TPMS-based architected materials. This underscores the need for a more general and systematic approach to enhance stiffness and isotropy through fundamentally different design principles simultaneously.

We propose a novel design framework for TPMS-based amorphousness-induced architected materials (TAAMs), which integrates controllable geometric randomness, termed designable amorphousness, into periodic TPMS architectures. Unlike stochastic foams or spinodal structures that rely on uncontrolled disorder [33, 34], TAAMs introduce tunable amorphous features through unit-cell-level modulation and rotation. This enables enhanced geometric diversity and mechanical tunability while preserving additive manufacturing compatibility [35,

36]. To explore the stiffness–isotropy trade-off in this expanded design space, we employ computational homogenization [33] and define effective stiffness and anisotropy metrics [26]. A multi-objective Bayesian optimization framework [37, 38] incorporating EHVI and PHVI strategies is used to identify Pareto-optimal TAAMs. The selected designs are fabricated via FDM [39] and validated through mechanical testing. As shown in later sections, TAAMs consistently outperform existing approaches [24, 27, 31, 32] in balancing stiffness and isotropy, demonstrating the robustness and generality of the proposed method.

The remainder of this manuscript is organized as follows. **Section 2** introduces the overall methodology of the proposed framework, including the generation of TAAMs, the computational homogenization process used to evaluate stiffness and anisotropy metrics, and the implementation of a multi-objective Bayesian optimization strategy. Two acquisition functions—Expected Hypervolume Improvement (EHVI) and Probability of Hypervolume Improvement (PHVI)—are applied to guide the optimization. **Section 3** presents the results of both numerical simulations and experimental validations. It compares the performance of optimized TAAMs with existing benchmark designs, demonstrating improvements in stiffness–isotropy trade-offs across various relative densities. Finally, **Section 4** summarizes the key findings, discusses the broader implications of designable amorphousness in architected materials, and outlines future research directions for expanding the TAAM framework to multifunctional and adaptive material systems.

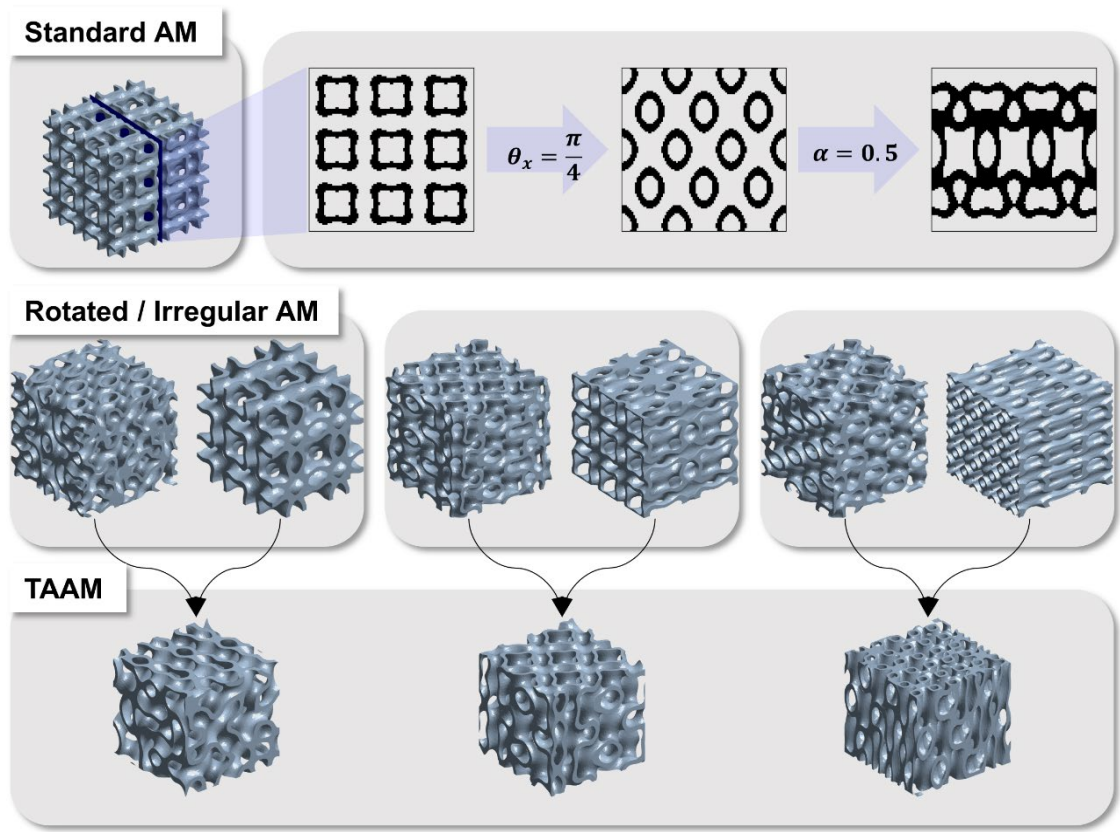

**Figure 1**. Schematic of design space construction for TAAMs. Top: transformation of regular TPMS structures via random rotation and amorphous modulation. Middle: intermediate configurations with either rotation or amorphousness applied. Bottom: final TAAMs combining both effects, exhibiting enhanced geometric diversity and improved isotropy potential.

## 2. Methodology

Achieving an optimal balance between stiffness and isotropy in TPMS-based architected materials has motivated the development of various advanced design strategies. This study introduces a novel framework for designing TPMS-based amorphousness-induced architected materials (TAAMs), which systematically address this trade-off by incorporating designable geometric disorder. The methodology consists of three main stages: data generation (**Section 2.1**), data-driven optimization (**Section 2.2**), and experimental validation (**Section 2.3**).

### 2.1. Data generation

To construct the TAAM design space, we initially selected Schoen's I-Wrapped Package (IWP) minimal surface as the base TPMS architecture. The rationale for this choice including a comparative analysis of anisotropy characteristics among various TPMS topologies, is provided in the **Supplementary Material** (**Figure S1** and **S2**).

#### 2.1.1. Design of TPMS-based amorphousness-induced architected materials

The geometry of TAAM was defined by the following implicit equation:

$$\phi_{IWP}(x, y, z) = 2(cosXcosY + cosYcosZ + cosZcosX) - (cos2X + cos2Y + cos2Z) = t$$

Here $X$, $Y$, and $Z$ are the scaled Cartesian coordinates given by:

$$X = \frac{2\alpha n_x \pi}{L_x} x, \ Y = \frac{2\beta n_y \pi}{L_y} y, \ Z = \frac{2\gamma n_z \pi}{L_z} z$$

In this formulation, $L_x$, $L_y$, and $L_z$ denote the dimensions of the unit cell along the $x$, $y$, and $z$ axes, respectively, while $t$ is the iso-value that determines the relative density of the structure.

The parameters $\alpha$, $\beta$, and $\gamma$ are defined as amorphousness constants that introduce geometric irregularity to enhance the overall isotropy of the structure. To ensure sufficient representativity of the amorphous design, $n_x = n_y = n_z = 3$ unit cells were used in each direction within the design domain. The actual implicit-based structure was generated by adopting a sheet-network type, considering the enclosed region satisfying $|\phi_{IWP}| < t$, with the relative density controlled by the iso-value $t$. To further introduce disorder, the orientation of each IWP unit cell was randomized using a rotation matrix $A$, defined as:

$$A = R_x(\theta_x) \cdot R_y(\theta_y) \cdot R_z(\theta_z)$$

where $R_x$, $R_y$, $R_z$ represent rotation matrices about the $x$, $y$, and $z$ axes, respectively. The rotated iso-surface is then given by: $\phi_{IWP}(r_x, r_y, r_z)$ where $[r_x \quad r_y \quad r_z]^T = A^{-1}[x \quad y \quad z]^T$.

In generating amorphous unit cells, the rotation angles $\theta_x$, $\theta_y$, and $\theta_z$ were randomly sampled from a uniform distribution within $-\pi$ to $\pi$, while the amorphousness constants $\alpha$, $\beta$, and $\gamma$ were sampled from 0.5 to 1.5. This disorder-assisted strategy, inspired by the approach described in the previous work [35], effectively expanded the design space and enhanced the likelihood of achieving isotropy.

### 2.1.2. Stiffness and anisotropy characterization via homogenization of cellular structures

The primary objective of this study is to achieve elastic isotropy without compromising stiffness. To quantify the anisotropic mechanical properties of TAAM, computational homogenization was employed to compute the effective stiffness tensor $\mathbb{C}$ [33]. The effective stiffness tensor represented as a fourth-order tensor satisfying the following symmetry conditions: $\mathbb{C}_{ijkl} = \mathbb{C}_{klij} = \mathbb{C}_{ijlk}$ for all $i,j,k,l = \{1,2,3\}$. Given the orthotropic nature of TAAM, the

stiffness tensor can be reduced to nine independent components:

$$\mathbb{C} = (\mathbb{C}_{1111}, \mathbb{C}_{1122}, \mathbb{C}_{1133}, \mathbb{C}_{2222}, \mathbb{C}_{2233}, \mathbb{C}_{3333}, \mathbb{C}_{2323}, \mathbb{C}_{3131}, \mathbb{C}_{1212})^T$$

This reduction simplifies the analysis significantly. The effective stiffness tensor for a given set of design parameters $\Theta = (\theta_x, \theta_y, \theta_z, \alpha, \beta, \gamma)$ was computed using computational homogenization via the finite element method (FEM). Each TAAM was embedded within a cubic domain, assuming an isotropic linear elastic constitutive law for the base material with a Young's modulus $E = 1$ and Poisson's ratio $\nu = 0.3$. Six loading conditions, including three axial and three shear cases, were applied under affine (linear) boundary conditions using an iterative implicit solver. For each loading case, the volume-averaged stress and strain were computed as $\bar{\sigma} = 1/|\Omega| \int \sigma \, d\Omega$ and $\bar{\varepsilon} = 1/|\Omega| \int \varepsilon \, d\Omega$, respectively.[40] where $\Omega$ denotes the cubic domain. The effective stiffness tensor was then obtained from Hooke's law: $\bar{\sigma} = \mathbb{C}:\bar{\varepsilon}$ [40].

To quantify anisotropy, the universal anisotropy index was adopted. Unlike Zener [41], Chung-Buessem [42], or Ledbetter-Migliori [43] indices, which are limited to cubic crystals or lack generality, universal anisotropy index [44] is applicable to non-cubic, non-periodic cellular structures. However, estimating the universal anisotropy index requires element-wise data, increasing computational cost [44]. To improve efficiency, anisotropy index was alternatively characterized using the ratio of the minimum to maximum directional elastic moduli: $E_{min}/E_{max}$. where $E_{min}$ and $E_{max}$ represent the minimum and maximum effective directional elastic moduli, respectively. The directional modulus $E(d)$ in direction $d$ was evaluated using [45, 46]:

$$\frac{1}{E(d)} = (d \otimes d):\mathbb{S}:(d \otimes d)$$

where $\mathbb{S}$ is the compliance tensor (inverse of $\mathbb{C}$). This directional modulus approach aligns well with universal anisotropy index, effectively capturing anisotropic behavior.

For benchmarking, the Hashin-Shtrikman (HS) upper bound [47] was employed as a reference for comparing the stiffness of different isotropic structures. The HS upper bound serves as an ideal measure for the maximum achievable stiffness of elastically isotropic cellular structures, providing a standard for evaluating how close a given design is to the theoretical maximum for isotropic configurations (anisotropy index = 1):

$$\frac{E}{E_s} = \frac{2\rho(5\nu_s - 7)}{13\rho + 12\nu_s - 2\rho\nu_s - 15\rho\nu_s^2 - 27}$$

where $E_s$ and $\nu_s$ denote the Young's modulus and Poisson's ratio of the base material, and $\rho$ is the relative density.

## 2.2. Data-driven optimization

To achieve an optimal balance between stiffness and isotropy, we employed a multi-objective Bayesian optimization (MBO) framework tailored for the design of TPMS-based amorphousness-induced architected materials (TAAM). This framework is built upon Gaussian Process Regression (GPR) as a surrogate model to approximate the mapping from design parameters to mechanical performance metrics.

Design candidates were sequentially updated using acquisition functions to guide the search toward improved solutions. In this study, we applied two different acquisition strategies—Expected Hypervolume Improvement (EHVI) and Probability of Hypervolume Improvement (PHVI)—to independently explore the design space. For each strategy, the optimization was conducted separately, and the final results were compared to evaluate how the acquisition function affects the performance of the resulting isotropic design.

### 2.2.1. Surrogate model construction

The first step in the MBO framework was the construction of the GPR-based surrogate model. GPR, as a probabilistic regression model, quantifies the uncertainty in the objective function $f(x)$ given the observation dataset $D = \{(x_i, y_i) | i = 1, \dots, n\}$. The covariance function was modeled using Matern 5/2 kernel, which effectively captures the complex input-output relationships inherent to the TAAM design space. The predictive distribution for a new design point $x^*$ was inferred from the multivariate Gaussian distribution, conditioned on the training data (observation data):

$$P_{y,y^*} = \begin{bmatrix} y \\ y^* \end{bmatrix} \sim \mathcal{N}\left(0, \begin{bmatrix} K & k \\ k^T & k(x^*, x^*) \end{bmatrix}\right)$$

where $K$ is the covariance matrix of the training data, $k$ is the covariance vector between the training data and the prediction point, and $k(x^*, x^*)$ is the self-covariance at $x^*$. The covariance function was defined as:

$$k(x_i, x_j) = \sigma_f^2 \left(1 + \frac{\sqrt{5}r}{l} + \frac{5r^2}{3l^2}\right) exp\left(-\frac{\sqrt{5}r}{l}\right) + \delta_{ij}\sigma_\varepsilon^2$$

where $r = \sqrt{(x_i - x_j)^T (x_i - x_j)}$, $\delta_{ij}$ denotes Kronecker delta function, and , $\sigma_f^2$, $l$, and $\sigma_\varepsilon^2$ are tunable hyperparameters. The optimal set of hyperparameters was determined through maximum likelihood estimation (MLE) using the L-BFGS optimizer for efficient convergence.

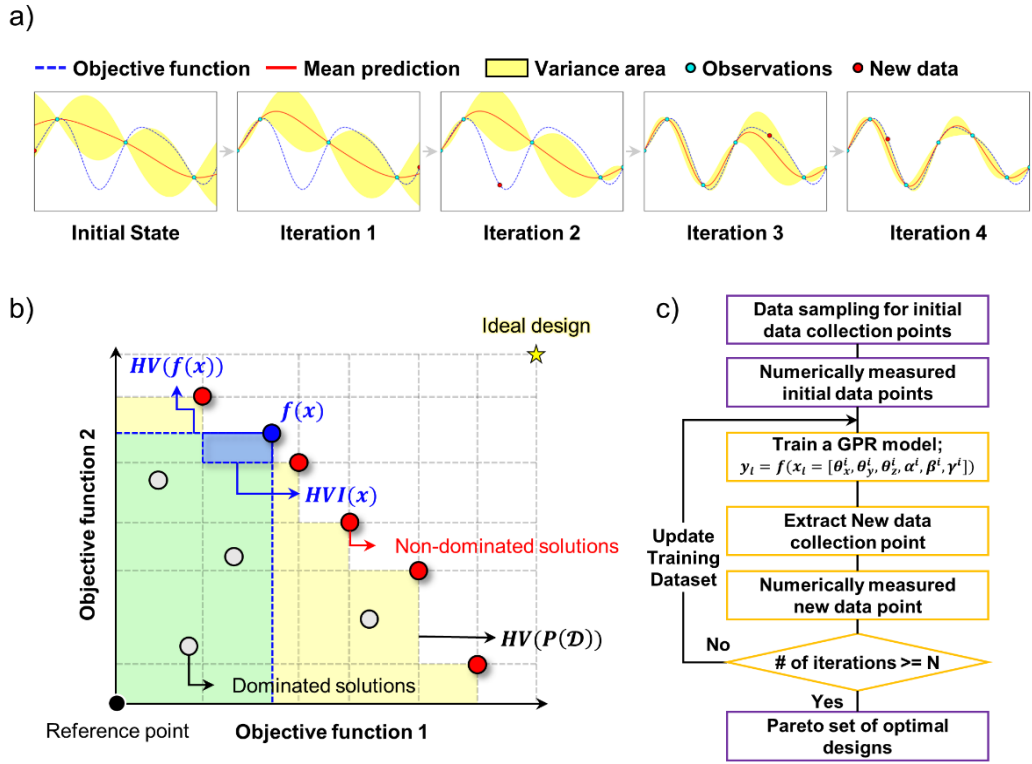

**Figure 2**. Multi-objective Bayesian optimization framework for TAAMs. (a) GPR surrogate modeling with uncertainty quantification. (b) Concept of hypervolume improvement and Pareto front expansion. (c) Overall optimization process integrating surrogate modeling, acquisition, and iterative updates.

**2.2.2. Acquisition strategies for multi-objective optimization**

To efficiently explore the design space, two acquisition functions were employed and comparatively analyzed: Expected Hypervolume Improvement (EHVI) and Probability of Hypervolume Improvement (PHVI). Both acquisition functions aim to guide the search process toward high-performing, Pareto-optimal designs, but differ in how they balance exploration and exploitation.

PHVI offers a probabilistic treatment of uncertainty, favoring exploitation by prioritizing candidates with a high likelihood of improving the Pareto front based on existing knowledge. This makes PHVI more effective in refining known high-performance regions. In contrast, EHVI focuses on the expected improvement in hypervolume and provides a better between exploration and exploitation, making it suitable for discovering diverse solutions in complex design spaces while still progressing toward optimal regions.

The initial dataset for Gaussian Process Regression (GPR) training was generated using Latin Hypercube Sampling (LHS) to ensure uniform coverage of the design space. A total of 20 design samples were used to construct the initial surrogate model. Subsequently, both EHVI and PHVI were applied in parallel optimization procedures, each guiding the selection of new design candidates over the course of 100 iterations. After each acquisition step, the selected designs were evaluated through finite element method (FEM) simulations, and the resulting data were added to the GPR training set.

Rather than combining results, each acquisition function was applied separately to assess its individual effectiveness in guiding the optimization process. This setup allowed for a fair comparison of the impact of acquisition function choice on the design trajectory and final mechanical performance. The comparative outcomes from EHVI- and PHVI-guided searches are discussed in the subsequent sections.

## 2.3. Experimental validation

### 2.3.1. Specimen fabrication via additive manufacturing

To validate the mechanical properties of the optimized TAAMs, physical specimens were fabricated using fused deposition modeling (FDM), a widely adopted additive manufacturing (AM) technique well-suited for constructing complex TPMS-based geometries without internal support structures [27]. This technique minimizes post-processing needs and ensures geometric fidelity, especially for intricate surfaces.

Printing was performed using a Prusa i4 MK4s system with PLA filament (Prusament PLA, Prusa Polymers, Czech Republic), chosen for its dimensional stability, low warpage, and compatibility with fine geometric features (see **Table 1**). A layer height of 0.1mm and nozzle width of 0.4mm were used, with an infill density of 100%. A rectilinear grid pattern was adopted with 90° rotation between adjacent layers to improve interlayer adhesion and mechanical consistency across directions [48].

**Table 1**. FDM-based additive manufacturing process parameters.

| Specification | Parameter |
| --- | --- |
| Machine | Prusa i4 MK4s |
| Material | PLA |
| Layer height | 0.1 (mm) |
| Nozzle extrusion width | 0.4 (mm) |
| Infill relative density | 100 (%) |
| Speed | 12 (mm/s) |

Each specimen was designed as a cube with a nominal edge length of 50mm and a relative density of approximately 30%, generated from the implicit surface function of the IWP-based TAAM geometry. Two representative structures were fabricated, each corresponding to a distinct optimization strategy: one from the PHVI-guided process and the other from EHVI-guided optimization. These were selected to examine the influence of different acquisition functions on stiffness and isotropy. Geometry generation was performed using a custom MATLAB-based framework, and STL files were converted to G-code using PrusaSlicer v2.6.1, which also managed toolpath alignment for enhanced surface quality.

The accuracy and reproducibility of the AM process were confirmed by measuring the actual relative densities of the printed samples. As listed in **Table 2**, the printed densities were consistent with the design targets, showing minor variations across principal directions: for PHVI-optimized TAAM, values ranged from 0.292 to 0.333; for the EHVI-optimized TAAM, from 0.279 to 0.297. These measurements validate the manufacturing consistency.

**Table 2**. Specimen relative densities for stl design file and as printed.

| Design | $\rho$, stl design | $\rho$, as printed | | |
|---|---|---|---|---|
| | | e1 | e2 | e3 |
| Optimized TAAM (PHVI) | 0.300 | 0.292 ± 0.001 | 0.333 ± 0.002 | 0.329 ± 0.007 |
| Optimized TAAM (EHVI) | | 0.297 ± 0.020 | 0.279 ± 0.123 | 0.290 ± 0.014 |

### 2.3.2. Mechanical testing protocol

Quasi-static uniaxial compression tests were conducted using a universal testing machine (Instron 5583) with a 100kN load cell. Each TAAM specimen was tested along its three principal axes (e1, e2, and e3), with the z-axis aligned to the printing direction. A displacement-controlled loading condition was applied at a constant strain rate of 1% per minute, and data were recorded at 50Hz.

To mitigate surface irregularities from the printing process and stabilize the initial stiffness response, a preconditioning step was implemented: five loading loading-unloading cycles up to 2% strain were applied prior to the main test. The final compression was performed up to densification, defined as a 70% reduction in initial specimen height.

Several key mechanical properties were quantitatively defined from the resulting stress-strain data. Crushing strength ($\sigma_{cr}$) was defined as the stress corresponding to the first prominent local maximum (initial peak) in the stress–strain curve, reflecting the initial load-bearing capacity before significant structural collapse or local buckling phenomena occurred. Plateau strength ($\sigma_{pl}$) was defined as the mean stress value within a specific strain range (0.2 to 0.3), characterizing the stable stress region during progressive structural collapse and plastic deformation. Energy absorption efficiency ($\eta$) was calculated as:

$$\eta(\varepsilon) = \frac{1}{\sigma(\varepsilon)} \int_0^\varepsilon \sigma(e) de$$

where $\sigma(\varepsilon)$ denotes the engineering stress at strain $\varepsilon$. The densification strain ($\varepsilon_d$) was identified as the strain at which the energy absorption efficiency ($\eta$) reached its maximum, marking the onset of substantial structural densification and indicating the termination of efficient energy absorption. Subsequently, energy absorption per unit volume ($W$) was computed by integrating the area under stress–strain curve from zero strain up to the determined densification strain ($\varepsilon_d$), quantifying the total energy dissipated by structural deformation. Young's modulus was extracted from the linear region (strain ≤ 2%) using linear regression.

Moreover, preliminary mechanical property validations were conducted through tensile and compressive tests, following ASTM D638 and ASTM D695 standards, respectively. The tensile test yielded a Young's modulus of 1357 ± 14 MPa, whereas the compressive test resulted in a modulus of 1630 ± 21 MPa. These results confirm the reliability of the selected additive manufacturing process and the suitability of the PLA material for fabricating geometrically complex TPMS-based structures.

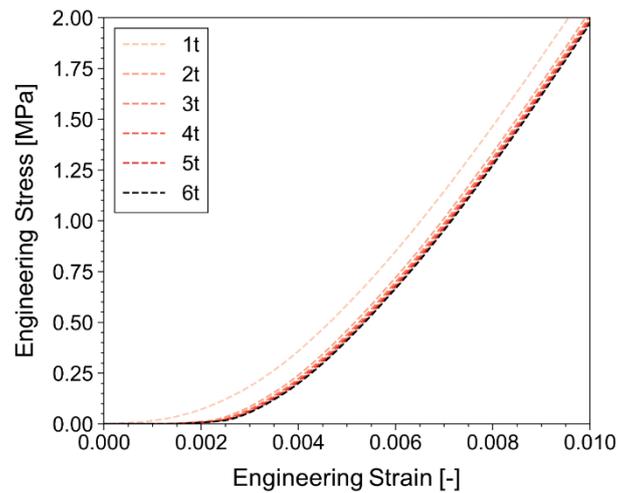

**Figure 3**. Repeated compression test results under small strain (≤ 2%) for TAAMs. The first cycle shows deviations due to surface irregularities, while subsequent cycles converge, validating stiffness measurements.

## 3. Results and Discussion

### 3.1. Numerical results

**Figure 4a** presents the Pareto fronts derived from the proposed multi-objective Bayesian optimization (MBO) using PHVI and EHVI acquisition functions. The results reveal a clear trade-off between the anisotropy index and the maximum directional elastic modulus. Compared to prior benchmark designs—including standard IWP, isotropic IWP from previous studies [24, 27], HTAM [31], and CFCC [32]—the proposed framework uncovers a broader spectrum of design solutions with significantly improved property trade-offs.

Unlike conventional design strategies that are often confined to a limited subset of the design space, the amorphousness-induced parameterization in this work enables the discovery of previously unexplored geometries that simultaneously achieve high stiffness and improved isotropy. This expanded design space facilitated the identification of novel configurations that offer a better balance between competing mechanical properties.

The design selected from the PHVI-guided optimization (**Figure 4b**) exhibits a slightly higher stiffness, whereas the EHVI-based design (**Figure 4c**) yields superior isotropic performance. The magnified view in **Figure 4d** highlights how these optimized TAAMs fill previously unoccupied regions of the stiffness-isotropy design space, underscoring the framework's capability to go beyond conventional TPMS limitations.

To verify the generalizability of the optimization framework, additional simulations were conducted at different relative densities ($\rho$ = 0.2 and $\rho$ = 0.4). As shown in **Figure S3** of the **Supplementary Material**, the proposed TAAM approach consistently achieved superior stiffness-isotropy trade-offs compared to conventional benchmarks, demonstrating robustness across a wide range of density conditions. Moreover, a comparison between the two acquisition strategies reveals that the EHVI-guided optimization trends to explore a broader region of the design space, offering diverse design candidates. In contrast, PHVI demonstrates a stronger capability in refining the Pareto front, leading to designs with slightly superior stiffness-isotropy balances. These trends were consistently observed across all tested relative densities, highlighting the complementary nature of the two acquisition functions and reinforcing the adaptability of the proposed framework.

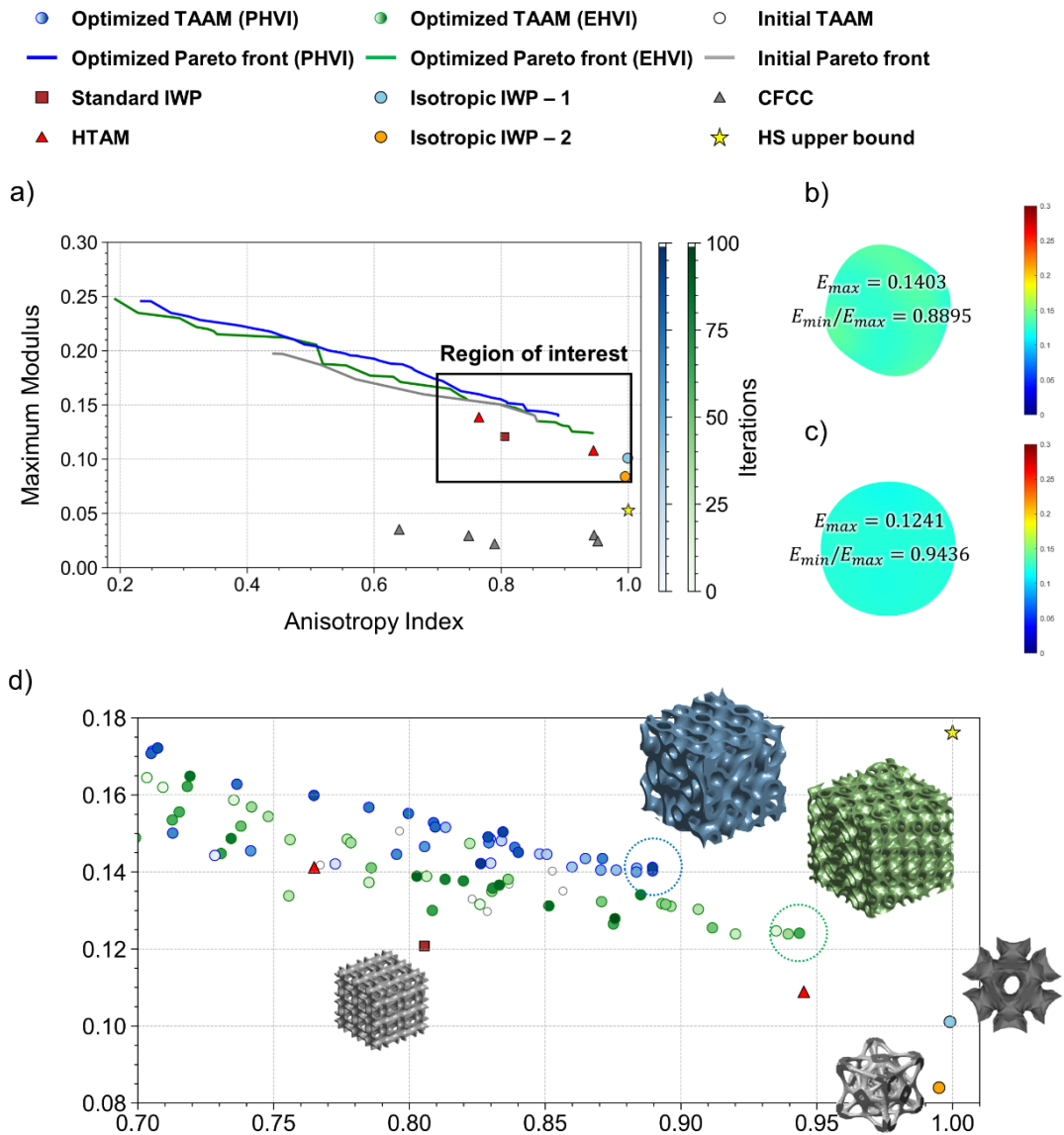

**Figure 4**. Optimization results and benchmark comparison. (a) Pareto fronts from PHVI and EHVI showing trade-off between stiffness and isotropy. (b, c) Directional modulus visualizations of PHVI- and EHVI-optimized TAAMs, respectively. (d) Close-up view of isotropic region with 3D morphologies of selected designs. Benchmark designs: orange circle [27], skyblue circle [24], red triangles [31], and grey triangles [32].

## 3.2. Experimental results

### 3.2.1. Linear elastic characteristics under small strain compression

As shown in **Figure 5**, the experimentally fabricated TAAMs demonstrate consistent elastic responses along all principal directions. The measured elastic moduli generally agree well with FEM predictions, with most relative errors falling below 5%, as summarized in Table 3. The largest deviation of 7.03% occurred along the e3 direction of the PHVI-optimized design, which may be attributed to local geometric irregularities introduced during additive manufacturing. This strong agreement validates the reliability of the optimization process and the accuracy of the fabrication method [49].

**Table 3**. Comparison between numerically predicted and experimentally measured elastic moduli of optimized TAAMs under small-strain loading.

| Design | Orientation | $E/E_s$ Numerical | $E/E_s$ Experimental | relative error [%] |
|---|---|---|---|---|
| Optimized TAAM (PHVI) | e1 | 0.1403 | 0.1448 ± 0.0003 | 3.23 |
| | e2 | 0.1321 | 0.1345 ± 0.0002 | 1.85 |
| | e3 | 0.1386 | 0.1483 ± 0.0006 | 7.03 |
| Optimized TAAM (EHVI) | e1 | 0.1230 | 0.1241 ± 0.0040 | 0.90 |
| | e2 | 0.1222 | 0.1202 ± 0.0074 | 1.65 |
| | e3 | 0.1170 | 0.1227 ± 0.0009 | 4.84 |

The EHVI-optimized TAAM, which emphasizes isotropy in the optimization objective, exhibited excellent directional uniformity in elastic moduli (0.1202–0.1241), confirming its near-isotropic behavior under small-strain loading. In comparison the PHVI-optimized TAAM yielded slightly higher stiffness values, especially along the e3 direction, but at the cost of increased directional variation. These results demonstrate that incorporating amorphousness and random unit cell rotations is effective in enhancing isotropy, and that the choice of acquisition function in MBO can guide the design towards specific mechanical targets such as stiffness or isotropy.

### 3.2.2. Nonlinear behaviors under large strain compressions

**Table 4** summarizes the nonlinear mechanical properties of the optimized TAAMs under large-strain compression. Both PHVI- and EHVI-optimized TAAM exhibited stable post-yield behavior, characterized by smooth stress-plateau regions and high energy absorption capacity, indicative of bending-dominated deformation modes.

**Table 4**. Nonlinear mechanical properties of optimized TAAMs under large-strain compression.

| Design | Orientation | $\sigma_{cr}$ [MPa] | $\sigma_{pl}$ [MPa] | $W$ [MPa] |
|---|---|---|---|---|
| Optimized TAAM (PHVI) | e1 | 18.408 ± 0.223 | 17.410 ± 0.010 | 8.916 ± 0.292 |
| | e2 | 15.990 ± 0.064 | 11.948 ± 0.199 | 7.509 ± 0.072 |
| | e3 | 19.500 ± 0.015 | 16.904 ± 0.189 | 9.510 ± 0.265 |
| Optimized TAAM (EHVI) | e1 | 15.449 ± 0.072 | 13.299 ± 0.032 | 7.695 ± 0.398 |
| | e2 | 15.930 ± 0.080 | 14.264 ± 0.072 | 8.924 ± 0.060 |
| | e3 | 14.932 ± 0.059 | 13.342 ± 0.030 | 8.478 ± 0.185 |

Notably, the EHVI-optimized TAAM showed exceptional isotropic behavior even in the nonlinear regime. As illustrated in **Figure 5b** (right), the engineering stress-strain curves along the three principal directions (e1, e2, and e3) exhibit near-perfect overlap up to densification. This strong directional agreement confirms that the proposed design strategy successfully preserves mechanical isotropy beyond the linear elastic range—an essential feature for applications involving multi-directional impact or crushing scenarios.

In contrast, the PHVI-optimized TAAM achieved higher peak strength (up to 19.5 MPa along e3) and greater total energy absorption (around 9.510 MPa) but displayed more pronounced directional variation, particularly in the early stages of plastic collapse. This highlights the intrinsic trade-off between maximizing stiffness and ensuring isotropy, as reflected in the contrasting optimization objectives.

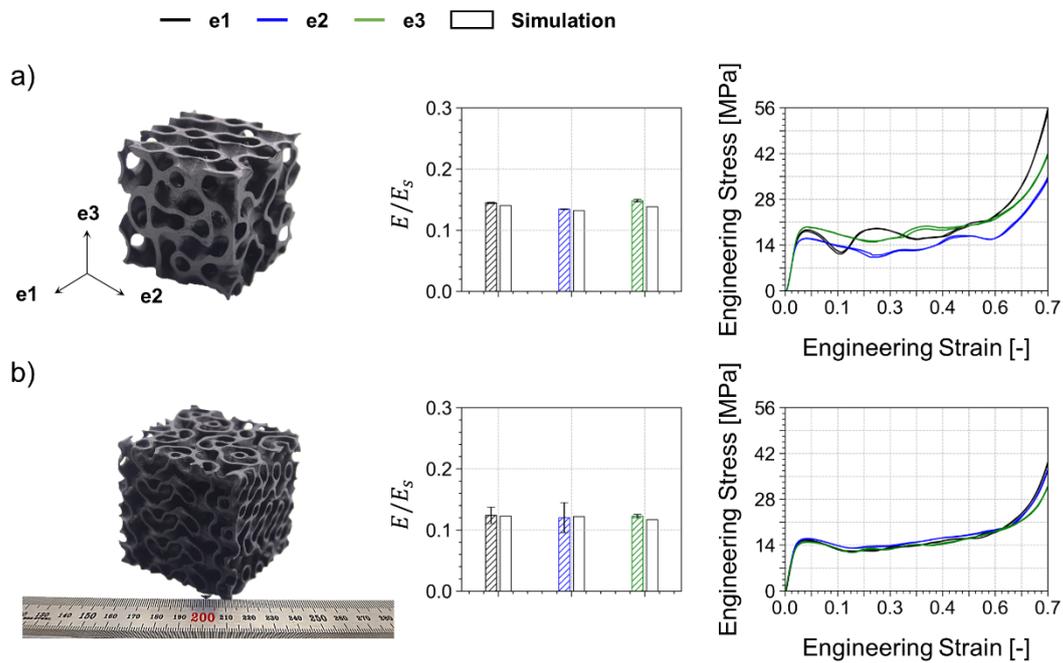

**Figure 5**. Experimental validation of (a) PHVI-optimized and (b) EHVI-optimized TAAMs. Left: fabricated specimens. Middle: comparison of simulated and measured elastic moduli along principal directions. Right: stress–strain curves showing consistent response, confirming isotropy in both linear and nonlinear regimes.

# 4. Conclusion

This study presented a data-driven framework for designing TPMS-based amorphousness-induced architected materials (TAAMs) to balance stiffness and elastic isotropy. We systematically expanded the conventional TPMS design space to include tunable geometric disorder as a controllable design parameter by introducing *designable amorphousness* through random unit cell rotation and geometric modulation. This new approach enables enhanced isotropy while preserving mechanical performance and manufacturability.

To efficiently navigate the expanded design space, we employed a multi-objective Bayesian optimization (MBO) framework using two acquisition strategies: Expected Hypervolume Improvement (EHVI) and Probability of Hypervolume Improvement (PHVI). These guided the discovery of TAAM configurations with distinct stiffness–isotropy trade-offs. Representative designs were fabricated using fused deposition modeling (FDM) and validated through quasi-static uniaxial compression tests. The experimental results showed strong agreement with numerical predictions, confirming both the effectiveness and feasibility of the proposed design framework. Compared to conventional TPMS benchmarks, the TAAMs demonstrated superior stiffness–isotropy trade-offs across various relative densities, highlighting the robustness and generalizability of the method.

Elastic isotropy is critical in structural applications subjected to unpredictable or multi-axial loading, such as biomedical implants, impact protection systems, and aerospace components. The proposed integration of designable amorphousness offers a paradigm shift in architected materials design, enabling the targeted introduction of local disorder to achieve globally isotropic behavior, unlike the incidental randomness found in stochastic structures. Looking ahead, the TAAM framework hold promise as a foundation for developing multifunctional architected materials that incorporate mechanical, thermal, acoustic, or biological functionalities. Its compatibility with AI-based generative design tools also opens pathways for integration into large-scale, application-specific optimization workflows. Moreover, by incorporating stimuli-responsive materials or active control schemes, TAAMs may evolve into programmable or adaptive metamaterials for real-time performance tuning in advanced engineering applications.

# Supplementary Material

# Data-driven Design of Isotropic and High-Stiffness TPMS-based Amorphousness-Induced Architected Material (TAAM)

Minwoo Park[a,+], Junheui Jo[a,+], and Seunghwa Ryu[a,*]

[a]Department of Mechanical Engineering, Korea Advanced Institute of Science and Technology, 291 Daehak-ro, Yuseong-Gu, Daejeon 34141, Republic of Korea

**Abstract**

For their excellent stiffness-to-weight characteristics, triply periodic minimal surfaces (TPMS) are widely adopted in architected materials. However, their geometric regularity often leads to elastic anisotropy, limiting their effectiveness under complex loading. To address this, we propose TPMS-based amorphousness-induced architected materials (TAAMs), which incorporate controllable geometric disorder as a tunable design variable. This concept of *designable amorphousness* broadens the geometric design space, enabling the simultaneous optimization of stiffness and isotropy. A data-driven framework integrating computational homogenization with multi-objective Bayesian optimization is employed to discover high-performance TAAMs. Selected designs were fabricated using fused deposition modeling and validated through uniaxial compression tests, showing strong agreement with numerical predictions. Compared to conventional TPMS, TAAMs exhibit significantly improved elastic isotropy while maintaining high stiffness across various relative densities. This approach offers a robust and scalable pathway for developing architected materials tailored to applications requiring isotropic performance, such as biomedical implants, protective systems, and aerospace components.

To justify the selection of Schoen's I-Wrapped Package (IWP) minimal surface as the primary base architecture in our TAAM framework, we conducted a comparative analysis across multiple conventional TPMS topologies including FRD, Gyroid, Diamond, Neovius, Primitive, and FKS.

As shown in **Figure S1**, which plots the ratio of maximum to minimum directional moduli $E_{max}/E_{min}$ over varying relative densities, it is evident that both the IWP and Primitive structures exhibit significantly higher anisotropy at low densities (< 0.3) which is the regime of primary interest in architected materials research due to the emphasis on lightweight design [1]. Among the surveyed topologies, the IWP shows the steepest increase in stiffness anisotropy as density decreases, making it an ideal candidate to test whether elastic isotropy can be recovered without sacrificing stiffness performance.

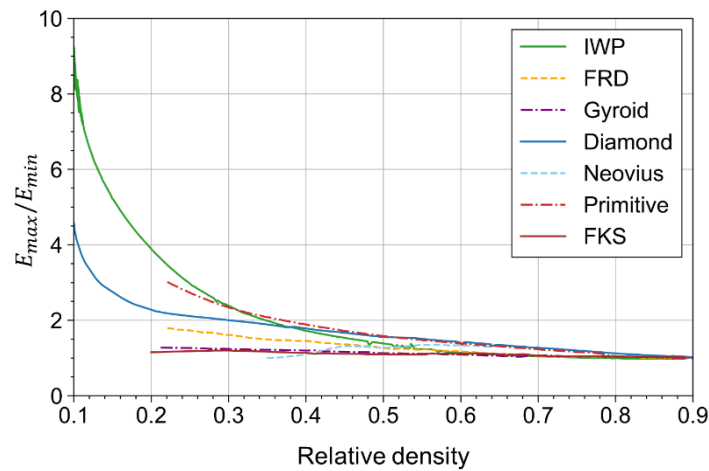

**Figure S1**. Comparison of anisotropy across various TPMS topologies (IWP, FRD, Gyroid, Diamond, Neovius, Primitive, and FKS) as a function of relative density.

**Figure S2** further demonstrates the effectiveness of the proposed TAAM design strategy by comparing the Pareto front progression under the same relative density ($\rho = 0.3$) for both IWP-based and Primitive-based TAAMs. Initial designs from both topologies show wide anisotropy distributions and modest stiffness. However, upon optimization via the MBO framework, the IWP-based TAAMs achieve a more favorable Pareto front, particularly in the low anisotropy region. This indicates that IWP-based TAAMs possess greater tunability under amorphousness modulation and random cell rotation, which allows a more pronounced enhancement in isotropy with minimal stiffness loss.

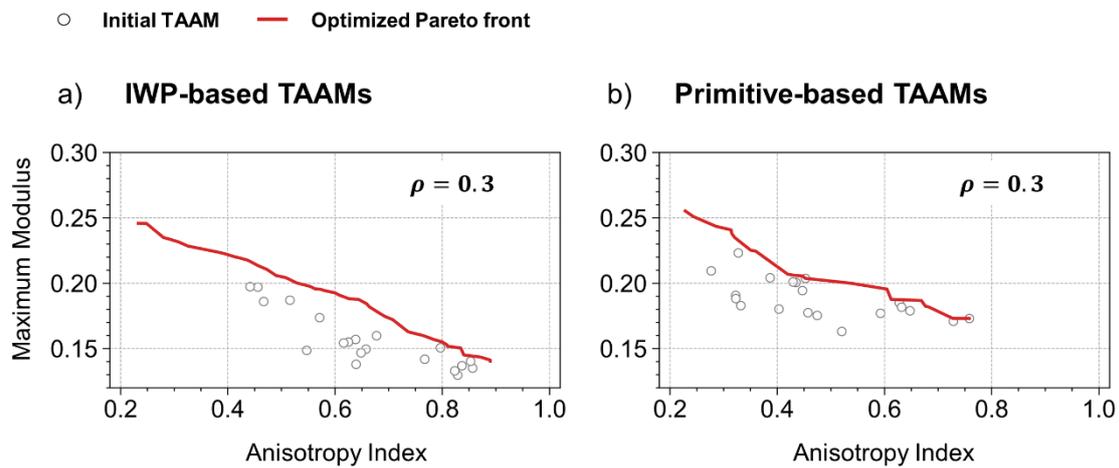

**Figure S2**. Pareto fronts of (a) IWP-based and (b) Primitive-based TAAMs at a fixed relative density ($\rho = 0.3$).

These results collectively support the central hypothesis of this study: starting from highly anisotropic TPMS architectures such as IWP enables the proposed TAAM framework to demonstrate its capacity to simultaneously achieve isotropy and high stiffness. Thus, focusing on the IWP topology not only highlights the strength of the method but also serves as a stringent benchmark to evaluate its generalizability and performance.

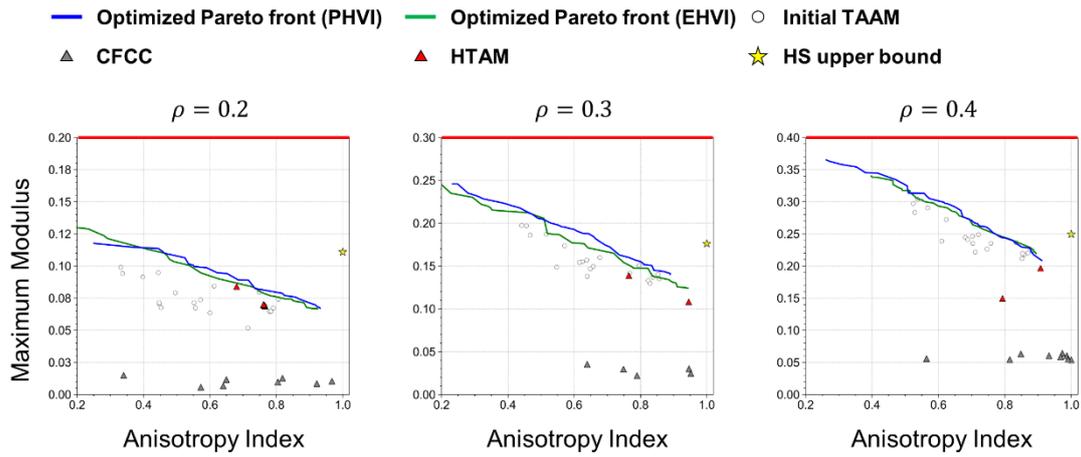

**Figure S3**. Comparison of TAAM designs and benchmark designs at varying relative densities ($\rho$ = 0.2, 0.3, 0.4) in terms of anisotropy index versus maximum directional elastic modulus. Blue and green lines represent the Pareto fronts obtained from PHVI- and EHVI-guided optimization, respectively. White circles denote the initial TAAM design candidates, while grey and red triangles represent CFCC [2] and HTAM [3] benchmarks from previous studies. Yellow stars indicate the theoretical Hashin-Shtrikman (HS) upper bound values under the assumption of perfect isotropy. The red lines mark the Voigt limit, corresponding to the theoretical maximum stiffness under the assumption of uniform strain. The proposed TAAM framework consistently outperforms conventional benchmarks by achieving a broader and superior balance between stiffness and isotropy across various density levels.